\newcommand\nodata{{...} }

\newcommand\etal{et~al.}
\newcommand\kms{\ifmmode {\rm\,km\,s^{-1}}\else${\rm\,km\,s^{-1}}$\fi}

\def\spose#1{\hbox to 0pt{#1\hss}}
\newcommand\simlt{\mathrel{\spose{\lower 3pt\hbox{$\mathchar"218$}}
     \raise 2.0pt\hbox{$\mathchar"13C$}}}
\newcommand\simgt{\mathrel{\spose{\lower 3pt\hbox{$\mathchar"218$}}
     \raise 2.0pt\hbox{$\mathchar"13E$}}}
\newcommand\aap{{\em A\&A}}
\newcommand\aaps{{\em A\&AS}}
\newcommand\aj{{\em AJ}}
\newcommand\apj{{\em ApJ}}
\newcommand\apjl{{\em ApJ}}
\newcommand\apjs{{\em ApJS}}

\newcommand\mnras{{\em MNRAS}}
\newcommand\nat{{\em Nature}}

\documentclass[usenatbib]{mn2e}

\usepackage[dvips]{graphicx}
\usepackage{psfig}
\usepackage{rotating}

\begin{document}

\title[Radiogalaxies in the SDSS]{Radiogalaxies in the Sloan Digital Sky Survey: spectral index-environment correlations}

\author[C. Bornancini \etal]{
\parbox[t]{\textwidth}{
Carlos~G.~Bornancini$^{1,2}$, Ana Laura O'Mill$^{1,2}$, Sebasti\'an Gurovich$^{1,2}$ and Diego Garc\'ia Lambas$^{1,2}$
}
\vspace*{6pt} \\
$^1$Instituto de Astronom\'\i a Te\'orica y Experimental, IATE, Observatorio Astron\'omico, Universidad Nacional de C\'ordoba,\\ 
Laprida 854, X5000BGR, C\'ordoba, Argentina.\\
$^2$Consejo Nacional de Investigaciones Cient\'\i ficas y T\'ecnicas (CONICET),
Avenida Rivadavia 1917, C1033AAJ, Buenos Aires, Argentina.\\
}
\pubyear{2010}

\maketitle

\begin{abstract}
We analyze optical and radio properties of radiogalaxies detected in
the Sloan Digital Sky Survey (SDSS). The sample of radio sources are
selected from the catalogue of Kimball \& Ivezi\'c (2008) with flux
densities at 325, 1400 and 4850 MHz, using WENSS, NVSS and GB6 radio
surveys and from flux measurements at 74 MHz taken from VLA
Low-frequency Sky Survey \citep{cohen}. We study radiogalaxy spectral
properties using radio colour-colour diagrams and find that our sample
follows a single power law from 74 to 4850 MHz. 
The spectral index vs. spectroscopic redshift relation ($\alpha-z$) is not significant for our sample of radio sources.
We analyze a subsample of radio sources associated with clusters of
galaxies identified from the maxBCG catalogue and find that about 40\%
of radio sources with ultra steep spectra (USS, $\alpha<-1$, where
$S_\nu \propto \nu^{\alpha}$) are associated with galaxy clusters or
groups of galaxies.  We construct a Hubble diagram of USS radio
sources in the optical $r$ band up to $z\sim0$.8 and compare our
results with those for normal galaxies selected from different optical
surveys and find that USS radio sources are around as luminous as
the central galaxies in the maxBCG cluster sample and typically more
than 4 magnitudes brighter than normal galaxies at $z\sim0$.3.

We study correlations between spectral index, richness and luminosity
of clusters associated with radio sources. We find that USS at low
redshift are rare, most of them reside in regions of unusually high ambient
density, such of those found in rich cluster of galaxies.  Our results
also suggest that clusters of galaxies associated with steeper than the average spectra have higher richness counts and are populated by
luminous galaxies in comparison with those environments associated to radio sources with flatter than the average spectra. A plausible explanation for our
results is that radio emission is more pressure confined in higher gas
density environments such as those found in rich clusters of galaxies
and as a consequence radio lobes in rich galaxy clusters will expand
adiabatically and lose energy via synchrotron and inverse Compton
losses, resulting in a steeper radio spectra.

\end{abstract}

\begin{keywords} 
surveys -- radio continuum: general -- radio continuum: galaxies --
galaxies: high-redshift
\end{keywords}

\section{Introduction}

Radio sources are frequently associated with massive systems at low and high redshifts. In the Local Universe these objects are usually identified with evolved red ellipticals and with luminous cD galaxies located at the centres of clusters of galaxies \citep{west}. Radio sources at high redshifts ($z\sim4$) are identified with massive forming systems, i.e galaxies with diffuse UV morphologies and consistent with small substructures around a dominant bright galaxy \citep{miley}.
 
Several works show that some distant radio sources are embedded in highly spatially extended ionized gas nebulae of 100-200 kpc \citep{vene02, villar}. These gas structures are observed in massive ellipticals or cD galaxies at the center of nearby clusters of galaxies. This evidence suggests that distant radio sources represent the progenitors of the most massive galaxies observed in the Local Universe and therefore important for the study of structure formation, such as clusters or groups of galaxies.

A high fraction of radio sources associated with clusters of galaxies have steep radio spectra as measured by the slope between two fixed ranges of frequencies, i.e the spectral index $\alpha$ (USS, $\alpha<-1$, where $S_{\nu} \propto  \nu^{\alpha}$, \citet{rot,cha96,jarvis01}).
Such a correlations between spectral index and redshift ($\alpha-z$ relation), has been used with success to search for distant galaxies \citep{DB00, jarvis04, cruz06, brode}.

There are at least three main explanations given in the literature for this phenomenon: The first is a radio k-correction. The spectral energy distribution (SED) of radio sources is usually concave, i.e the spectral index increases with frequency. The observed spectral index determined from two fixed observed frequencies will therefore sample a steeper part of the radio spectrum for sources at higher redshifts. The second is related with the interaction between photons of the cosmic microwave background radiation (CMB) and the relativistic electron population observed in the plasma gas of radio sources. The energy density of the CMB increases as $(1+z)^4$ and hence Inverse Compton scattering (IC) of the CMB becomes increasingly important for sources at high redshifts \citep{krolik}. The third is related to an intrinsic correlation between radio luminosity and {\it rest-frame} spectral index \citep{blun, cham} that due to a Malmquist like bias in flux density limited surveys translates into a correlation between spectral index and redshift. An alternative explanation is related with an ambient density effect. The presence of USS radio sources residing closest to cluster centres may be due to a manifestation of pressure-confined radio lobes which slow adiabatic expansion of the plasma. Radio lobes will be pressure-confined and lose energy primarily via synchrotron and IC losses \citep{klamer}. 

Studies of radio sources at low and moderate redshifts ($z <0.5$) show different environments. 

Hill \& Lilly (1991) find that only 50\% of powerful radio sources at $z\sim0.5$ are located in rich galaxy clusters, even though similar sources avoid such environments at low redshifts. However,  Geach et al. (2007) analyzed radio sources in the Subaru/XMM-Newton Deep Field and found that low-power radio galaxies at $z\sim0.5$ reside in moderately rich groups - intermediate environments between poor groups and rich clusters.

\citet{prestage} investigate the local galaxy density around powerful radio sources using the angular cross-correlation technique and find that compact radio sources appear to lie in regions of low galaxy density. Moreover, that complex Fanaroff-Riley class I sources are typically found in regions of significantly enhanced galaxy density.

\citet{alli} study the evolution of galaxies in radio-selected groups at $z<0.5$ with the same range of radio power and find that strong radio galaxies are located in a wide range of environments but not as wide as for groups in general. At low redshifts ($z<0.1$) radio-selected groups have the same richness and blue fraction as do optically-selected groups. At high redshifts ($z\sim 0.4$) all groups have the same proportion of blue galaxies. 

Studying the prevalence of radio loud AGN activity in nearby groups and galaxy clusters, selected from the SDSS catalog, \citet{best} find that brightest group and cluster galaxies are more likely to host a radio loud AGN than other galaxies of the same stellar mass.

In this work we analyze optical and radio properties of radio sources with steep radio spectra, and we determine the main characteristics of clusters of galaxies associated with radio sources. 

The structure of this paper is organized as follows: Section 2
describes the optical and radio samples analyzed.  We analyze radio
spectral properties of our sample in Section 3.  Section 4 compares
the optical and the radio luminosities of sources associated with
steep spectrum and central galaxy clusters. In section 5, we study
the Hubble diagram in the optical $r$ band for USS sources. We study
in Section 6 spectral index and richness properties of galaxies
associated with clusters of galaxies and finally, in Section 7, we
discuss our main results.

Throughout this work we assume a standard $\Lambda$CDM model Universe with cosmological parameters, $\Omega_{M}$=0.3, $\Omega_{\Lambda}$=0.7 and a Hubble constant of $H_0=$100~h~km~s$^{-1}$Mpc$^{-1}$.

\section{Radio and optical Galaxy samples}

The sample of radio sources was selected from the catalogue of \citet{kimball}. This catalogue presents information in radio and optical bands taken from several surveys, including Faint Images of the Radio Sky survey (FIRST, 1400 MHz), NRAO VLA Sky Survey (NVSS, 1400 MHz), Westerbork Northern Sky Survey (WENSS, 325 MHz), Green Bank survey (GB6, 4850 MHz ), and the Sloan Digital Sky Survey (SDSS) optical survey. The flux density limits for each radio catalogue are 1, 2.5, 18 and 18 mJy for the FIRST, NVSS, GB and WENSS surveys, respectively.

The SDSS catalogue contains flux densities of detected objects measured nearly simultaneously in u, g, r, i, and z optical bands (Fukugita et al. 1996) with a limiting magnitude of $r < 22.2$ in an area of  $\sim 10^4$ square degrees in the North Galactic cap and a small region of 225 square degrees in the South Galactic cap.

The sample of optical galaxies in the fields of radio sources were selected using the Catalog Archive Server Jobs System (CASJOBS) interface of SDSS, which allows one to obtain catalogues with parameters from the SDSS survey (DR6). \footnote{http://casjobs.sdss.org/CasJobs/}.

The matching radius to correlate the radio (FIRST) and optical (SDSS) surveys was 2$\arcsec$. 
We chose this matching radius in order to avoid contamination by line of sight matches of physical unrelated objects, since \citet{kimball} previously found an efficiency (fraction of matches which are physically real) of 95\% and completeness (fraction of real matches that were found) of 98\% with this matching radius.
In this work we used the following catalogue subsets: Sample D (detected by FIRST, NVSS and WENSS survey, with 63,660 sources), sample E (detected by FIRST, NVSS, WENSS, GB6 and the SDSS survey, 4732 sources) and sample G (detected by FIRST, NVSS, WENSS and the SDSS spectroscopic survey, 2885 sources).
For all these samples \citet{kimball} estimated matching efficiencies of $> 80$\% and completeness limits of $> 90$\%, using an appropriate matching radius. 

A more detailed description of the radio sample and the criteria used to merge catalogues can be found in \citet{kimball}. 

Our sample of tracer galaxies for this work are drawn from the catalogue of galaxies with photometric redshifts of the DR6 catalogue \citep{oya}.
These photometric redshifts were calculated using a Artificial Neural Network technique and the Nearest Neighbor Error (NNE) method to estimate photo-z errors for 77 million objects classified as galaxies in DR6 with $r < 22$. In order to study and analyze the properties of clusters of galaxies associated with radio sources, we correlate the radio sources detected from the FIRST, NVSS and WENSS catalogues with clusters of galaxies selected from the maxBCG survey \citep{max} using a matching radius of $10\arcsec$.
The maxBCG catalogue includes 13,823 clusters selected from the SDSS photometric data, using two well-known features of rich clusters of galaxies: the red-sequence observed as a ridge line in the colour-magnitude diagram (the E/S0 ridgeline) and the presence of a bright dominant central galaxy (BCG). The catalogue includes the position, cluster photometric redshifts, BCG spectroscopic redshifts, $r$ e $i$ total galaxy luminosities and luminosities of the central galaxies, the cluster richness, $N_{\rm gal}$ calculated as the number of galaxies projected within 1 $h^{-1}$ Mpc, brighter than $0.4L_{*}$ and with colours matching the E/S0 ridgeline, the cluster richness within $R_{200}$ from the cluster center, defined as the mean density of 200 times the mean density of the Universe (see for instance \citet{maxII}).  

\begin{figure}
\includegraphics[width=8cm]{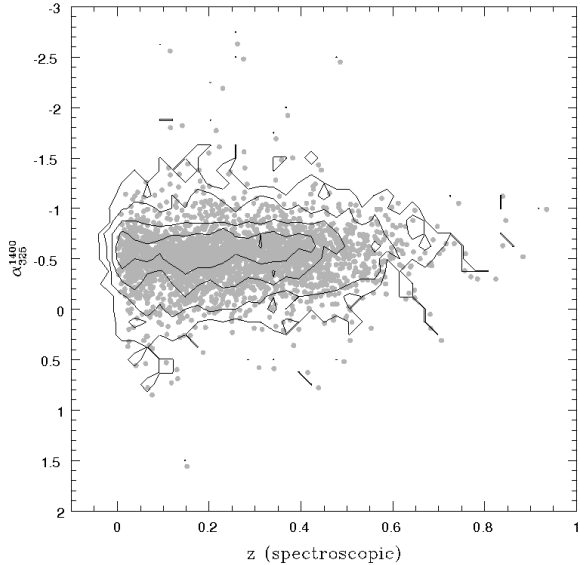}
\caption{Spectral index, $\alpha_{325}^{1400}$, vs. spectroscopic redshift from the SDSS catalogue for radio sources detected in FIRST (1.4 GHz), NVSS (1.4 GHz), WENSS (325 MHz) and the SDSS spectroscopic catalogue with a matching radius of 2$\arcsec$. The contours represent the 50, 70, 80 and 90\% of total objects. } 
\label{az}
\end{figure}

\section{Radio Properties}

\subsection{$\alpha-z$ relation}

In Figure \ref{az} we plot spectral index $\alpha_{325}^{1400}$ vs. spectroscopic redshift for radio sources identified in the SDSS catalogue. The data were obtained from the catalogue subset G taken from \citet{kimball}. 
The contours represent the 50, 70, 80 and 90\% of total objects. We do not find a tendency of $\alpha$ with redshift, showing that the most distant ($z > 0$.5) radio galaxies in our sample do not show steep spectra due to a k-correction effect applied to a concave radio SED.

\subsection{Analysis of the SED in radio frequencies}

\begin{figure}
\includegraphics[width=9cm]{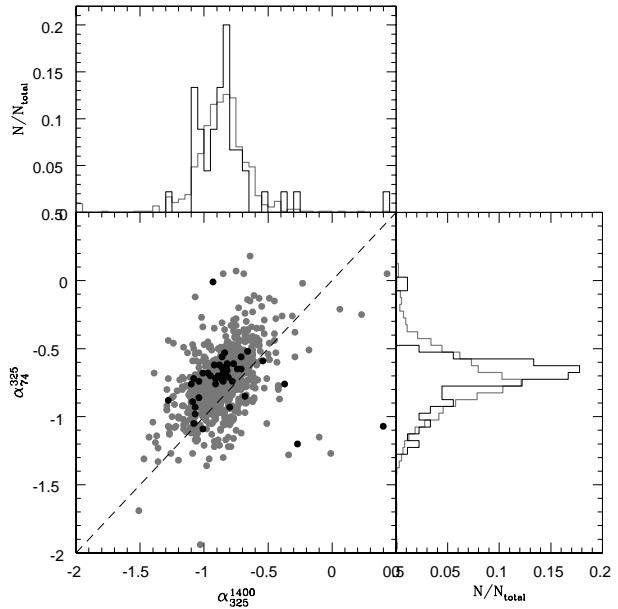}
\caption{Radio Colour-colour diagram distributions for sources in the VLSS (74 MHz), WENSS (325 MHz) and NVSS (1.4 GHz) samples (grey circles). Filled circles represent measurements obtained for radio sources associated with maxBCG clusters. The dashed line indicates the relation for radio sources whose spectra follow a single power law from 74 to 1400 MHz.Histograms represent the corresponding distribution of spectral index.}
\label{74}
\end{figure}

\begin{figure}
\includegraphics[width=9cm]{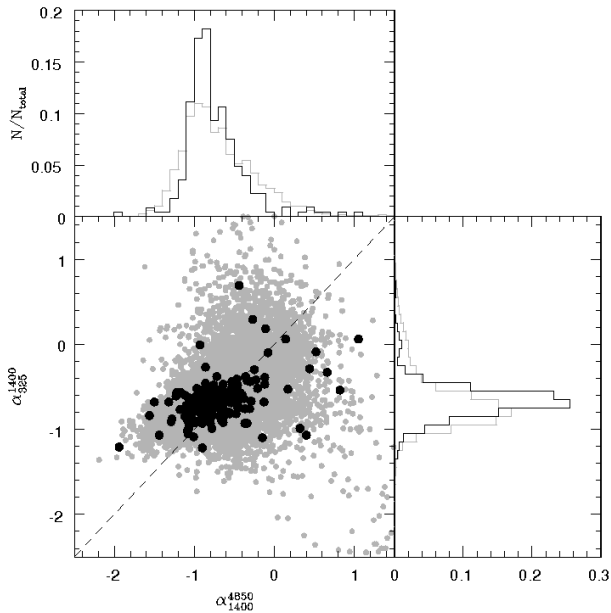}
\caption{Radio Colour-colour diagram distributions for sources in the WENSS (325 MHz), NVSS (1.4 GHz) and GB6 (4.85 GHz) samples (grey circles). Filled circles represent measurements obtained for radio sources associated with maxBCG clusters. The dashed lines correspond to a single power law spectra from 325 to 4850 MHz.Histograms represent the corresponding distribution of spectral index.}
\label{gb6}
\end{figure}

In order to study radiogalaxy properties we analyze radio colour-colour diagrams. 
In Figure \ref{74} we show the radio two-colour diagram which compares the spectral index at 74-352 MHz and 352-1400 MHz. We use the CATS database of the Special Astronomy Observatory \citep{verkho} in order to obtain flux measurements at 74 MHz taken from the VLA Low-frequency Sky Survey (VLSS) \citep{cohen}, using a matching radius of 60$\arcsec$ for the FIRST and NVSS samples.  Grey circles represent sources detected in the FIRST, NVSS and WENSS catalogues (listed as Sample D in \citet{kimball}). Filled circles represent data for radio sources associated with clusters of galaxies identified in the maxBCG catalogue.
The dashed line indicates the relation for radio sources whose spectra follow a single power law from 74 to 1400 MHz. Again, we do not find a clear tendency that most of radio sources show some flattening towards low frequencies.
In Figure \ref{gb6} we show a similar analysis for radio sources with radio emission in 325, 1400 and 4850 MHz, listed as Sample E in \citet{kimball} (grey circles). Filled circles represent data obtained for radio sources identified with clusters of galaxies detected in the maxBCG catalogue. The equal number of points on either side of this line indicate no significant spectral curvature.

\section{Optical and radio luminosities} 

In order to study radio source luminosity properties at optical and radio frequencies as well as other possible correlations, we calculated the {\it rest-frame} radio luminosity at 1.4 GHz:

\begin{equation}
\centering
L_{1.4}=4\pi~{D^{2}_{L}(z)}S_{1.4}~(1+z)^{-(1+\alpha)},
\end{equation}
where  $D_{L}(z)$ is the luminosity distance in the adopted $\Lambda$--CDM cosmology, $S_{1.4}$ is the observed flux density at 1.4 GHz and $(1+z)^{-(1+\alpha)}$ is the standard k-correction term used in radio frequencies. 

\begin{figure}
\includegraphics[width=8cm]{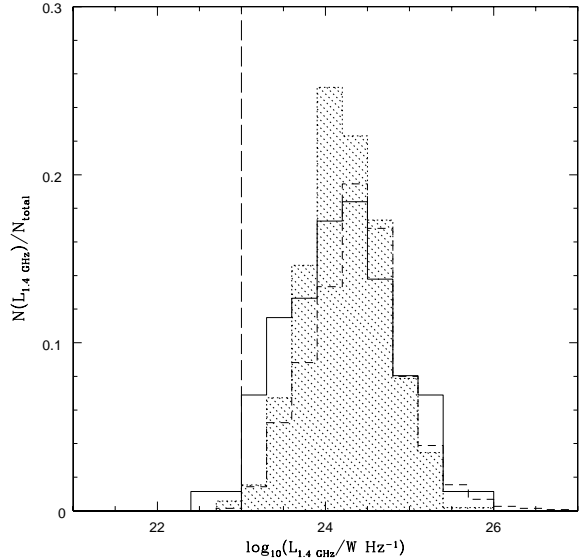}
\caption{Radio luminosity distribution in 1.4 GHz. Shaded histogram represents the luminosity distribution for maxBCG clusters detected in the FIRST, NVSS and WENSS survey. Dashed line histogram shows the distribution for radio sources without a maxBCG cluster association. Solid line histogram represents the distribution for USS sources. Dashed line shows the criterion for radio--loud radio sources ($L_{1.4 GHz}>10^{23}$ W Hz$^{-1}$)} 
\label{rl}
\end{figure}

In Figure \ref{rl} we show the luminosity distribution of radio
sources associated with maxBCG clusters (shaded histogram) and USS
($\alpha_{325}^{1400} < -1$) radio sources with spectroscopic
redshifts (solid line histogram). The dashed line histogram shows the
distribution for radio sources without a maxBCG cluster
association. We note the good agreement between these three
distributions, indicating that most of the radio sources associated
with central clusters of galaxies and USS sources are radio-loud
($L_{1.4 GHz}>10^{23}$ W Hz$^{-1}$). Several works show a correlation
between spectral index and radio luminosity (or radio power)
\citep[$e.g.,$][]{blunb, gopal, laing}.  As can be seen, we do not
find any trend in radio luminosity and spectral index for these
subsamples of radio sources.  It should be noted that the results
shown in this figure are not likely to be biased by possible
luminosity selection effects in steep spectral index sources.

\begin{figure}
\includegraphics[width=8cm]{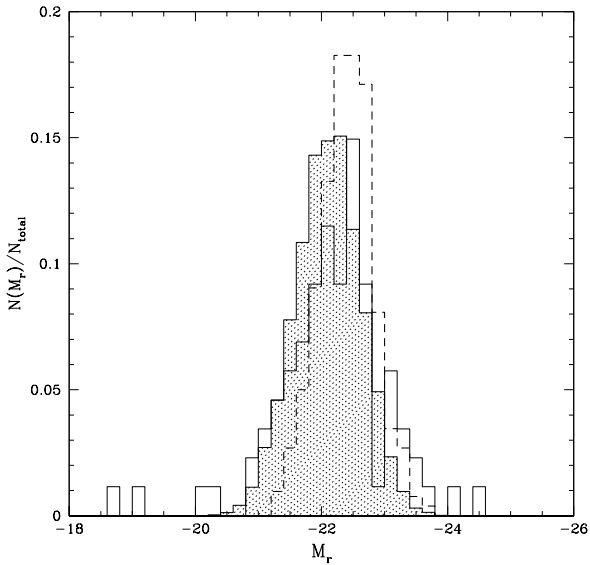}
\caption{Absolute magnitude distribution in $r$ band. Shaded histogram represents the distribution for central galaxies in the maxBCG cluster catalogue. Solid and dashed line histograms shows the corresponding distribution for USS sources and for radio sources associated with clusters of galaxies identified in the maxBCG catalogue, respectively.}  
\label{Mr}
\end{figure}

To obtain absolute magnitudes in the $r$ band we use the de-reddened
model magnitudes and k-correct using the V4.1 public code of
\citet{blanton}.  Figure \ref{Mr} shows the distribution of absolute
$r$ band magnitude for USS radio sources (solid line histogram) and
for central cluster objects in the maxBCG catalogue (shaded line
histogram). The dashed line histogram shows the distribution obtained
for radio sources associated with maxBCG clusters. As can be seen
there is only a small difference between the two samples, where USS
radio sources are marginally brighter than the central galaxies associated with
maxBCG clusters.

\section{Hubble diagram in the {$\rm r$} band}

In the last three decades the Hubble diagram in the $K-$ band ($K-z$
diagram) has been used to detect and study distant radio galaxies. One
of the first works using the Hubble diagram was publish by
\citet{lilly84} using a sample of radio sources detected in the 3CR
catalogue. The first radiogalaxy identified at $z > 3$ in fact was selected at
radio frequencies as a faint source in the $K-$band \citep{lilly88}.
This technique was used in combination with spectral index cuts and
rejecting radio sources with large angular sizes ($< 30\arcsec$)
\citep{DB00, debreuck02b, cohen04, cruz06}.

\begin{figure}
\includegraphics[width=8cm]{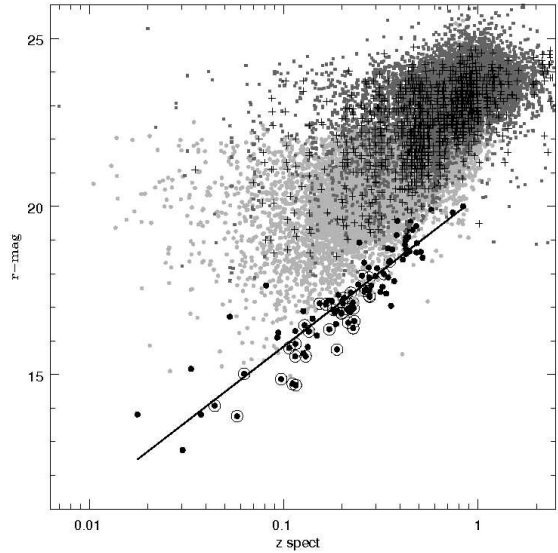}
\caption{Hubble diagram in the $r$ band vs. spectroscopic redshift taken from the SDSS catalogue. Filled circles represent measurements obtained for USS objects. USS radio sources identified with maxBCG clusters are displayed as big circles. Light grey circles correspond to galaxies detected in the SDSS catalogue in the NOAO Deep Wide--Field Survey area with photometric redshifts taken from \citep{oya}.
Dark grey points represent measurements of galaxies detected in the VIMOS VLT deep survey (VVDS-DEEP) (Le Fevre et al. 2005) with spectroscopic redshift and magnitudes converted to the SDSS system. Crosses represent spectroscopic measurements for galaxies identified in the Hubble Deep Field North  region obtained in the ACS-GOODS survey (Cowie et al. 2004). The solid line represent the best fit line obtained for the USS source sample}
\label{kz}
\end{figure}

\begin{figure}
\includegraphics[width=8cm]{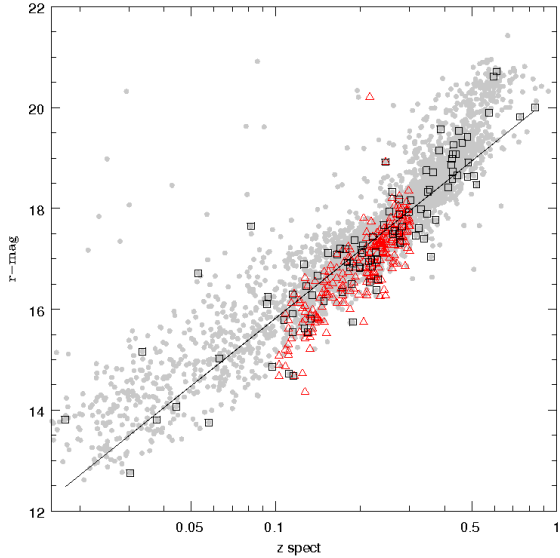}
\caption{Hubble diagram in the $r$ band vs. redshift. Radio sources with spectroscopic redshift measurements (grey points). Open red triangles represent measurements obtained for maxBCG clusters with radio emission. USS radio sources are displayed with open squares.}
\label{kz2}
\end{figure}

In this work we perform a similar analysis in the optical $r$
band. Figure \ref{kz} shows the Hubble diagram using the $r$ band
(petrosian magnitudes) vs. spectroscopic redshift. Filled circles
represent measurements obtained for USS objects, big circles represent
radio sources associated to maxBCG clusters. Light grey circles
correspond to galaxies detected in the SDSS catalogue within the NOAO
Deep Wide--Field Survey area, with photometric redshifts taken from
\citep{oya}. Dark grey points represent measurements of galaxies
detected in the VIMOS VLT deep survey (VVDS-DEEP) (Le Fevre et
al. 2005), with spectroscopic redshift and magnitudes converted to the
SDSS
system \footnote{http://www.sdss.org/dr4/algorithms/sdssUBVRITransform.html}. Crosses
represent spectroscopic measurements for galaxies identified in the
Hubble Deep Field North region obtained in the ACS-GOODS survey (Cowie
et al. 2004). The solid line represent the best fit obtained for the
USS source sample,
$r=(4.46\pm0.21)\times$log$_{10}$(z)$+(20.28\pm0.16)$.  As can be
seen, USS radio sources are typically more than 4 magnitudes brighter than
normal galaxies at $z\sim0$.3.  In Figure~\ref{kz2} we plot a similar
Hubble diagram for radio sources with spectroscopic redshift quoted in
sample G from \citet{kimball} (grey points). Open red triangles
represent measurements obtained for maxBCG clusters with radio
emission and USS radio sources are displayed with open squares.  The
solid line represents the best fit obtained for the USS source
sample. USS radio sources follow a similar distribution in comparison
with that obtained in the maxBCG cluster sample in the redshift range
$0.1<z<0.3$. We find that USS sources are typically as luminous as the
central galaxies in the maxBCG cluster sample.  In the next Section,
we analyze the nature of USS sources detected at low redshifts
($z<0.1$).

\section{Ultra steep spectrum sources}

In order to study the nature of radio sources with ultra steep spectra
at low redshifts, we select a set of galaxies detected by NVSS, FIRST,
WENSS, and SDSS with spectroscopic redshifts (catalogue subset G in
\citet{kimball}). We use the NED
database \footnote{http://nedwww.ipac.caltech.edu/ the NASA-IPAC
  Extragalactic Database.} to search for known objects present in the
literature and find that 40\% of USS sources are associated with
clusters or groups of galaxies identified in the maxBCG \citep{max},
Abell \citep{abell}, Zwicky \citep{zwi} catalogues and Northern Sky
optical Cluster Survey (NSCS, \citet{lopes}) catalogues. Table 1 lists
the sample of USS sources identified with galaxies with optical
spectra in the SDSS catalogue. The columns are the following: ID taken
from \citet{kimball}, position, spectroscopic redshift, spectral index
obtained between 352MHz and 1.4 GHz, 1.4 GHz radio luminosity, $r$
band absolute magnitude, and ID taken from literature.  In Figure
\ref{overlays} we show a sample of colour images of radio sources with
$\alpha_{325}^{1400} < -1$. At low redshift ($z<0.1$) we find that
most USS sources are associated with nearby bright spiral galaxies or
interacting systems, with a possible AGN. This may be due to the low
limiting flux density used ($S_{1400}=2.5$ mJy, from the NVSS survey)
since we note that \citet{DB00} found that they can select against low
redshift spiral galaxies among USS sources by selecting sources with
$S_{1400} >$ 10 mJy.

Some radio sources that are associated to bright red galaxies are
located in similar density environments as those found in clusters or
groups of galaxies, but do not have a cluster association in the
literature. In the next Section we analyze the galaxy density
associated with these systems.

\section{Spectral index and richness} 

\begin{figure}
\includegraphics[width=8cm]{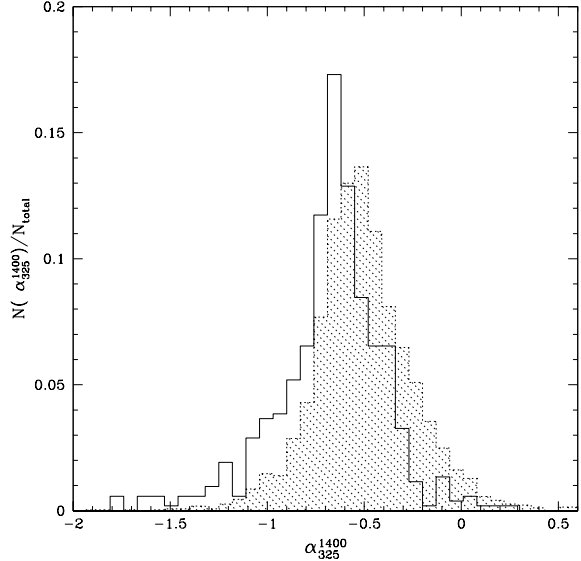}
\caption{Spectral index $\alpha_{325}^{1400}$ distribution for radio sources without a maxBCG match (shaded histogram)  and with a maxBCG cluster match (solid lines).}
\label{alfa}
\end{figure}

In order to study the nature of radio sources associated to clusters
of galaxies we analyze possible correlations between spectral index,
the richness associated with each cluster and the luminosity
properties of galaxies in these environments.  The spectral index
($\alpha_{325}^{1400}$) distribution for radio sources detected in
FIRST, NVSS and WENSS catalogues (listed as Sample D) not associated
with a maxBCG cluster can be seen in Figure \ref{alfa} (shaded
histogram). The solid line histogram represents the corresponding
spectral index distribution for radio sources associated with maxBCG
clusters.  The mean value distribution for clusters is
$\overline\alpha_{325}^{1400}=-0.65$.  We find a tendency that most
radio sources associated with maxBCG clusters have steep radio spectra
in comparison with field radio sources at similar redshifts. This
result is in agreement with \citet{baldwin} and \citet{slingo}, who
also find that nearby radio sources with steep spectra reside in rich
clusters of galaxies.

\begin{figure}
\includegraphics[width=8cm]{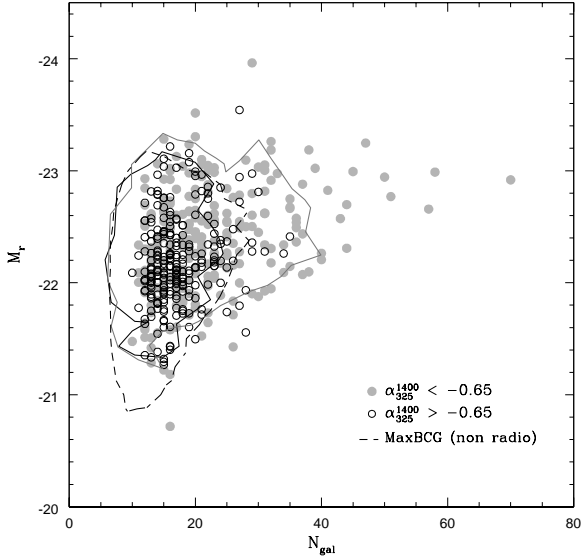}
\caption{Absolute magnitude in the $r$ band of central maxBCG cluster
  galaxies as a function of galaxy richness ($N_{gal}$ ). Grey circles
  represent measurements for clusters associated with radio sources
  with $\alpha_{325}^{1400}< -0.65$ and black circles represent
  clusters identified with sources with $\alpha_{325}^{1400}>
  -0.65$. The contours represent the 90\% of total objects (grey line
  for steep sources, solid line for flat sources and in dashed lines
  for maxBCG clusters without radio emission).}
\label{iso}
\end{figure}

In Figure \ref{iso} we plot the absolute $r$ band magnitude of central
maxBCG cluster galaxies vs. richness ($N_{\rm gal}$ ) for each cluster
associated to radio sources with steeper than average,
$\alpha_{325}^{1400}< -0.65$ (filled grey circles) and with flatter
than average spectra, $\alpha_{325}^{1400}> -0.65$ (open circles),
respectively. The contours represent the 90\% of total objects (grey
line) for steep sources, solid line for flat sources and dashed lines
for maxBCG clusters without radio emission. We find that clusters of
galaxies associated with steep spectrum sources have brighter central
galaxies and have a high galaxy richness in comparison with clusters
associated with flatter than the average radio sources.

\begin{figure}
\includegraphics[width=8cm]{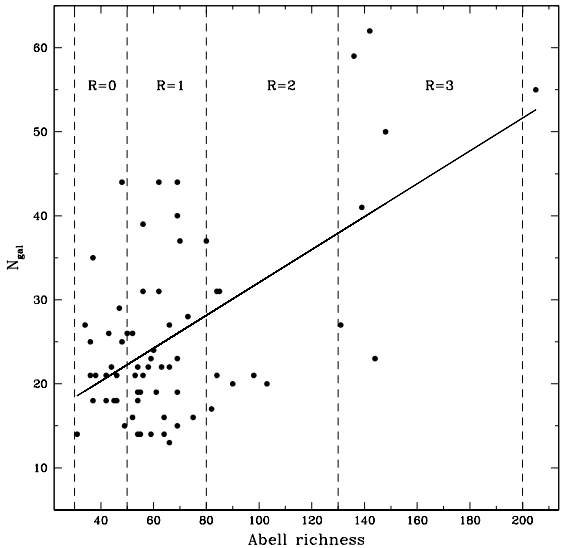}
\caption{Abell richness vs. maxBCG richness $N_{gal}$  from \citep{max}.
 Dashed lines shows the corresponding Abell richness group. Solid line represents the best fit obtained for these parameters.
}
\label{9a}
\end{figure}

In order to provide a suitable calibration to help the reader judge how 
Abell richness counts relate to the $N_{\rm gal}$ values, we show in  
Figure \ref{9a} this correlation using the revised northern 
Abell Catalog \citep{abell}.
As it can be appreciated in this figure, although with a large  
scatter, there is a positive correlation between these richness  
estimates. Abell richness class $R=0$ clusters have commonly $N_{\rm gal}$ $\sim$ 20, while objects with richness class 2 correspond to $N_{\rm gal}$$>$ 30. 

\begin{figure}
\includegraphics[width=8cm]{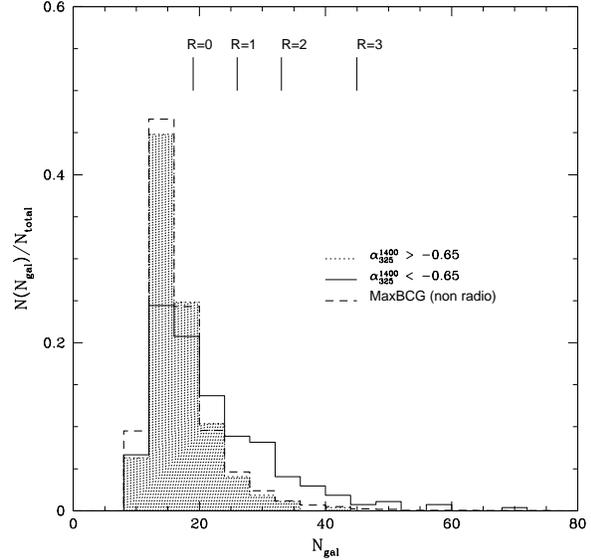}
\caption{Galaxy richness distribution for maxBCG clusters  in the WENSS--NVSS catalogues. The solid line histogram represents the richness associated to radio sources with $\alpha_{325}^{1400}< -0.65$ and the shaded histogram corresponds to sources with $\alpha_{325}^{1400}> -0.65$. The dashed line histogram shows the distribution for maxBCG clusters without radio emission.
The corresponding Abell richness class are marked at the top of the figure.}
\label{max2}
\end{figure}

In Figure \ref{max2} we present the distribution of richness ($N_{\rm gal}$ ) for clusters of galaxies taken from the maxBCG catalogue calculated from the red-sequence of the colour--magnitude diagram. We plot in solid line the richness distribution of radio sources with $\alpha_{325}^{1400}< -0.65$. The shaded histogram represents the distribution for radio sources with $\alpha_{325}^{1400}> -0.65$ and the dashed line histogram the richness distribution for maxBCG clusters without radio emission.
 We find a tendency that radio sources with steeper than the average spectra are found preferentially in higher galaxy richness environments than are radio sources with flatter than the average spectra.

For reference, we have also indicated in this figure the mean values of $N_{\rm gal}$ corresponding Abell richness class 0, 1, 2 and 3.

\begin{figure}
\includegraphics[width=8cm]{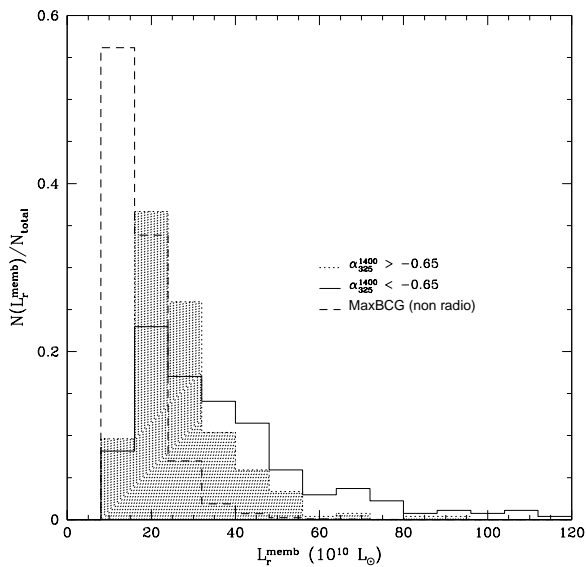}
\caption{Total luminosity distribution for galaxies associated with maxBCG clusters detected in the WENSS--NVSS catalogues. Solid line histogram corresponds to radio sources with $\alpha_{325}^{1400}< -0.65$ and the shaded histogram represents the corresponding distribution for sources with $\alpha_{325}^{1400}> -0.65$. The dashed line histogram shows the distribution obtained for maxBCG clusters without radio emission.}
\label{lumen}
\end{figure}

In order to compare the $N_{\rm gal}$  richness to other cluster parameters, we note the relation  $N_{\rm gal}$  and $R_{200}$ ($R_{200}$ $\sim$ $N_{gals}^{0.6}$)  \citep[See figure 7,][]{hansen}

In Figure \ref{lumen} we plot the total $r$ band luminosity of
galaxies for clusters of galaxies associated with radio sources with
$\alpha_{325}^{1400}< -0.65$ (solid line histogram) and those with
($\alpha_{325}^{1400}> -0.65$) as a shaded histogram. The dashed line
histogram represents the distribution obtained for maxBCG clusters
without radio emission. We find that radio sources with steeper than
average spectra in central clusters of galaxies are populated by
luminous galaxies in comparison with radio sources with flatter than
average spectra. In contrast clusters of galaxies without radio
emission have a lower distribution of total galaxy luminosity.

\begin{figure*}
\includegraphics[width=8cm]{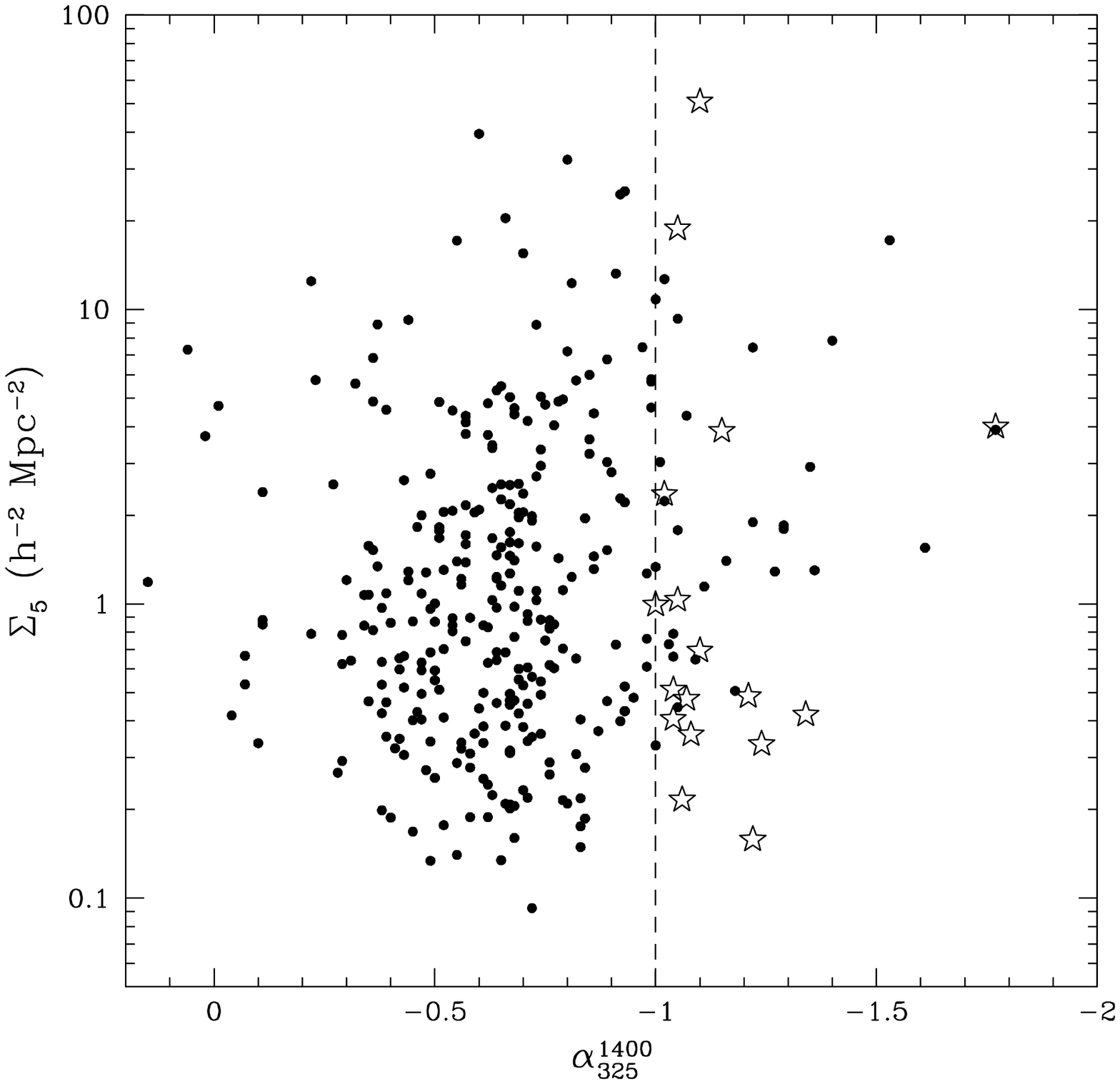}
\includegraphics[width=8cm]{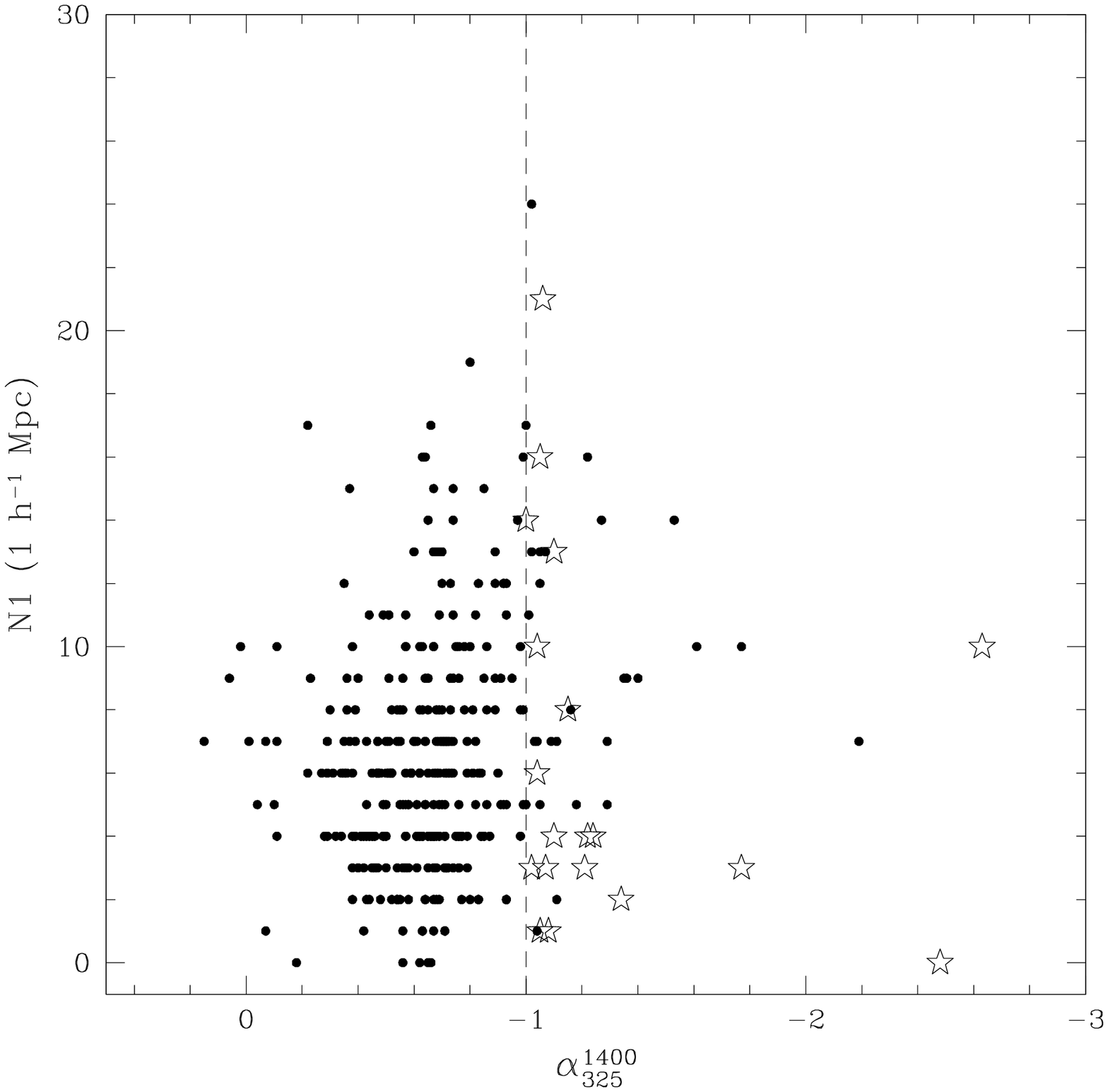}\\
\caption{Left panel: Logarithm of the 5$^{th}$ nearest neighbour local galaxy density estimator ($\Sigma_5$) as a function of the spectral index for maxBCG clusters. Dashed vertical line show the selection criterion for USS sources. Right panel: Projected density of galaxies within $< 1~h^{-1}$ Mpc vs. spectral index $\alpha_{325}^{1400}$ for cluster of galaxies in the maxBCG catalogue (filled circled). The projected density was calculated as a number of galaxies with photometric redshifts in the range $\Delta$z=0.01 from the BCG in the cluster in the redshift range $0.2< z <0.3$.  Stars represent values obtained for the USS sample without a maxBCG cluster  association.}
\label{sig5}
\end{figure*}

\begin{figure*}
\includegraphics[width=15cm]{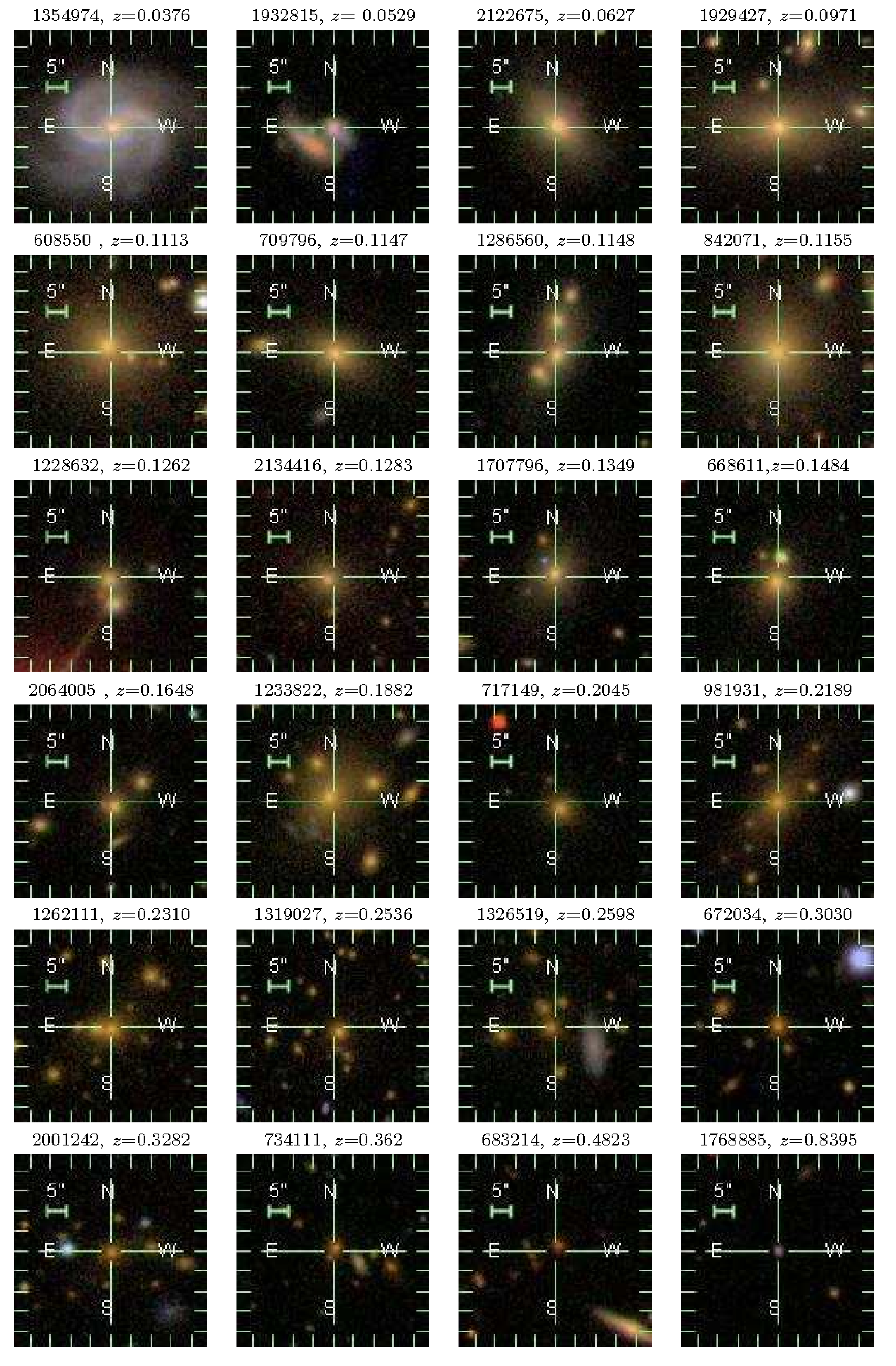}
\caption{Cutout colour images of  a subsample of galaxies associated with Ultra Steep radio sources in the SDSS catalogue. The open cross indicates the radio position taken from the FIRST survey.
The ID and redshift (increasing from left to right) are indicated above each plot. Information for the complete sample is listed in Table~1.} 
\label{overlays}
\end{figure*}

In a similar way, we analyze the density of galaxies associated with
different types of radio sources.  In Figure \ref{sig5} left panel, we
show the projected galaxy density using the 5$^{th}$ nearest neighbour
in the plane of sky ($\Sigma_5$, \citet{omill}) as a function of
spectral index for maxBCG clusters with $0.2< z <0.3$. The dashed line
shows the USS criterion.  Galaxies were selected with a range of
radial velocity difference ($\Delta$V), adopting a fixed
$\Delta$V=1000 km~s$^{-1}$ (for $z < 0.3$) and a varying
$\Delta$V=1000-3000 km~s$^{-1}$.

In Figure \ref{sig5} right panel, we plot the projected density of galaxies within 1~$h^{-1}$ Mpc and with photometric redshifts in the range $\Delta$z=0.01 from the central cluster galaxies (N1) as a function of the spectral index for clusters in the redshift range $0.2< z <0.3$ (filled circles). Stars represent the same values obtained for USS radio sources without a know cluster association, from the literature.
As can be seen in both plots, we found that these USS sources inhabit environments with galaxy densities similar to those clusters selected from the maxBCG catalogue.



\section{Conclusions}
We study optical and radio properties of radiogalaxies detected in the
Sloan Digital Sky Survey (SDSS) with flux densities of 74, 325, 1400
and 4850 MHz, using the VLSS, WENSS, NVSS and GB6 radio catalogues. We
search for a possible empirical correlation between the spectral index and
redshift, however we find no significant trend. We analyze the functional form of the SED using
colour-colour diagrams at radio frequencies.  We do not find a clear
tendency of radio sources to show flattening towards low frequencies,
as expected assuming concave curvature in the radio SED.
It is well known that a narrow relation exists between $K-$band and
redshift as observed in Hubble diagrams \citep{debreuck02b,
  willott,jarvis01, eales}.  In this work we construct a Hubble
diagram of USS radio sources in the optical $r$ band to
$z\sim0$.8. Despite any k-correction and possible extinction effects,
our $r$ band Hubble diagram (Figure 6) also clearly shows a tight
correlation.  We find that USS radio sources are as bright as central
galaxies in the maxBCG cluster sample and are typically more than 4
magnitudes brighter than normal galaxies at $z\sim0$.3. 
We note that this result is not entirely new, for example \citet{debreuck02b} also 
find that at redshifts $<\sim1$ radio-loud galaxies define the luminous envelope 
using near infrared $K-$band magnitudes.   

Regarding the possible dependence of radio luminosity on environment, we notice that the radio luminosity distribution of USS, radio sources in general, and radio sources in clusters are remarkably similar (Figure 4), indicating that USS prefer higher denstity environments, independent of radio luminosity.These results are consistent with those by \citet[fig. 8,][]{hill} who find no significant correlation in cluster richness $N_{0.5}$ and the rest-frame 2.0 GHz radio power for a sample of radio sources with $z<0.$5. Similar results were obtained by \citet{alli} who find no trend of richness with radio luminosity at 408 MHz for a sample of radiosources with $z<0.$5.

We also analyze the richness and spectral index properties of clusters
of galaxies associated with radio sources and find that 40\% of USS
sources identified in the SDSS spectroscopic catalogue are associated
with cluster or groups of galaxies identified in the literature, such
as in the maxBCG, Abell, or Zwicky catalogues. We analyze the local
density of galaxies around the sample of USS sources without a know
cluster association from the literature, using the $\Sigma_5$ and N1
estimators and find that these USS sources have similar galaxy
densities to clusters selected from the maxBCG catalogue.  

We also find that USS sources at low redshift are rare objects (99
from a total sample of 2885 radio sources detected in the SDSS
spectroscopic catalogue). However a majority reside in regions of
unusually high ambient density, such as those regions found in rich
cluster of galaxies.
 
Our results complement those found by \citet{DB00}. These
authors define a sample of 669 USS sources selected from the WENSS,
TEXAS, MRC, NVSS and PMN radio surveys.  They conclude that the
majority of relative nearby ($z<\sim 0.4$) USS objects are located in
galaxy clusters.  They find that at least 85\% of the X-ray objects
associates with USS sources are galaxy clusters or known groups from
the literature.

At lower redshifts, we find that radio sources with $\alpha_{325}^{1400}< -0.65$, 
are preferentially located in galaxy
cluster environments. 
This result contrast with \citet[fig. 7,][]{prestage} where it is found no dependence of  
the spatial cross--correlation amplitude on spectral index. We note although that this 
statistical analysis concerns more the large-scale, rather than the local environment of our study. 

We also find that clusters hosting radio sources with spectra steeper than 
the average have a higher galaxy richness and
are populated by brighter galaxies in comparison to clusters associated 
to radio sources with $\alpha_{325}^{1400}> -0.65$. A natural explanation for these
correlations is that radio emission in rich cluster of galaxies is
pressure-confined in a high gas density environment.  Radio lobes in
galaxy cluster environment will expand adiabatically and lose energy
via synchrotron and inverse Compton losses, resulting in a steeper
radio spectra \citep{klamer}.


\clearpage

\small
\begin{table*}
{\bf Table 1.} Sample of USS radio sources in the SDSS catalogue with spectroscopic redshifts.\\
\begin{center}
\begin{small}
\begin{tabular}{rrrrrcrc}
\hline
ID & R.A$^{radio}_{J2000}$ & DEC$^{radio}_{J2000}$ & z & $\alpha_{352}^{1400}$ & Log($L_{1.4}$)& $M_r$  & ID from literature\\
& $^h\;\; ^m\;\;\;\; ^s\;\;\,$ & \degr$\;\;\;$ \arcmin$\;\;\;$ \arcsec$\;$ &&&  W Hz$^{-1}$&&Designation\\
\hline
608550 &07 25 57.08&$+$41 23 05.13& 0.1113& $-$1.40& 23.76 & $-$22.18 &{\tt MaxBCG J111.48808+41.38519}\\
664059 &07 51 31.86&$+$43 49 29.49& 0.4249& $-$1.00& 24.22 & $-$22.63 & \nodata                    \\
668611 &07 53 32.47&$+$38 57 52.71& 0.1484& $-$1.04& 23.37 & $-$21.71 &  \nodata\\
672034 &07 54 57.66&$+$38 15 22.71& 0.3030& $-$1.10& 23.73 & $-$20.00 &  \nodata\\
683214 &07 59 49.48&$+$35 32 33.82& 0.4823& $-$1.04& 25.19 & $-$22.54 & \nodata\\
709796 &08 10 54.66&$+$49 11 03.90& 0.1147& $-$1.24& 23.28 & $-$22.27 &{\tt MaxBCG J122.72750+49.18436}\\
717149 &08 13 50.80&$+$39 32 32.11& 0.2045& $-$1.03& 23.90 & $-$21.79 &{\tt MaxBCG J123.46125+39.54183}\\
734111 &08 20 32.39&$+$30 34 48.65& 0.3628& $-$1.04& 25.98 & $-$21.97 & \nodata\\
735423 &08 21 03.64&$+$52 44 35.82& 0.4441& $-$1.38& 25.08 & $-$22.79 & \nodata\\
748224 &08 26 00.38&$+$40 58 51.75& 0.0576& $-$1.11& 22.78 & $-$22.79 &{\tt SDSS-C4-DR3 3247}\\
780016 &08 38 23.27&$+$29 45 21.67& 0.1068& $-$1.06& 22.68 & $-$22.00 & {\tt SHK 182 GGroup}\\
794634 &08 43 59.18&$+$51 05 25.55& 0.1264& $-$1.12& 24.37 & $-$22.63 & \nodata\\
801800 &08 46 37.85&$+$51 27 16.56& 0.1800& $-$1.27& 23.88 & $-$21.71 &{\tt MaxBCG J131.65796+51.45436}\\ 
814626 &08 51 17.29&$+$37 04 29.00& 0.2207& $-$1.00& 24.11 & $-$20.32 &\nodata\\
837935 &08 59 57.32&$+$56 47 12.15& 0.1833& $-$1.06& 23.71 & $-$22.11 & \nodata\\
838957 &09 00 20.28&$+$52 29 39.73& 0.0302& $-$1.02& 22.01 & $-$24.49 & {\tt CGCG 264-047}\\
842071 &09 01 30.10&$+$55 39 16.42& 0.1155& $-$1.80& 23.81 & $-$22.75 & {\tt MaxBCG J135.37558+55.65463 }\\  
846087 &09 03 00.14&$+$35 27 04.82& 0.3488& $-$1.01& 25.53 & $-$21.09 & \nodata\\
848084 &09 03 44.85&$+$41 38 19.31& 0.2189& $-$1.01& 24.42 & $-$21.42 & {\tt MaxBCG J135.93682+41.63908}\\ 
932725 &09 34 42.25&$+$35 14 16.48& 0.4618& $-$1.15& 24.31 & $-$22.62 & \nodata\\    
940538 &09 37 37.11&$+$37 05 35.37& 0.4492& $-$1.01& 25.24 & $-$22.53 & \nodata\\
955678 &09 43 09.29&$+$29 50 18.31& 0.2969& $-$1.05& 24.42 & $-$18.71 & \nodata \\
981931 &09 52 49.14&$+$51 53 04.99& 0.2151& $-$1.77& 24.08 & $-$22.49 &  {\tt ZwCl 0949.6+5207}\\
989139 &09 55 29.87&$+$60 23 17.47& 0.1989& $-$1.55& 23.82 & $-$22.03 & {\tt MaxBCG J148.87452+60.38814}  \\            
1002970&10 00 31.01&$+$44 08 42.94& 0.1533& $-$1.44& 23.16 & $-$21.32 & {\tt RBS 0819}               \\                      
1027752&10 09 28.28&$+$46 17 37.17& 0.3858& $-$1.04& 24.29 & $-$19.86      &             \\                       
1051000&10 17 58.13&$+$37 10 54.02& 0.0442& $-$1.16& 22.04 & $-$22.34 & {\tt CGCG 183-009 }         \\                       
1075981&10 27 09.98&$+$39 08 06.04& 0.3378& $-$1.02& 24.93 & $-$23.20 &  \nodata                   \\                       
1080596&10 28 54.68&$+$48 09 38.20& 0.4851& $-$1.45& 24.99 & $-$22.19 &  \nodata                   \\                      
1109542&10 39 32.11&$+$46 12 05.54& 0.1864& $-$1.21& 24.85 & $-$21.92 &  \nodata                  \\                       
1128337&10 46 25.51&$+$59 37 37.59& 0.2282& $-$1.07& 23.56 & $-$22.91 & {\tt MaxBCG J161.60635+59.62690} \\ 
1138526&10 50 10.03&$+$32 22 05.09& 0.1150& $-$1.33& 23.01 & $-$21.94 &  {\tt NSCS J105005+322256 /CLtr   }                \\ 
1160615&10 58 19.46&$+$41 03 40.76& 0.1299& $-$1.07& 23.41 & $-$22.44 & {\tt MaxBCG J164.58100+41.06140   }              \\  
1177112&11 04 33.11&$+$46 42 25.96& 0.1410& $-$1.82& 23.45 & $-$21.08 &  \nodata            				\\  
1179743&11 05 30.73&$+$31 14 36.74& 0.4381& $-$1.43& 24.12 & $-$21.34 &  \nodata           				\\  
1216462&11 18 45.25&$+$52 16 00.95& 0.4309& $-$1.24& 24.60 & $-$22.18 &  \nodata           				\\  
1228632&11 23 22.90&$+$47 55 14.34& 0.1262& $-$1.03& 22.91 & $-$21.08 &  \nodata          				\\  
1233822&11 25 16.31&$+$42 29 10.97& 0.1882& $-$1.02& 24.12 & $-$23.00 & {\tt ABELL 1253  }                              \\  
1239404&11 27 18.46&$+$53 02 21.12& 0.3236& $-$1.04& 24.49 & $-$23.56 &  \nodata       					\\  
1244341&11 29 01.60&$+$32 45 50.65& 0.5759& $-$1.10& 25.02 & $-$22.24 &  \nodata            				\\  
1250737&11 31 20.94&$+$33 34 46.95& 0.2219& $-$1.61& 23.72 & $-$22.26 & {\tt MaxBCG J172.83707+33.57975}\\ 
1260844&11 34 57.39&$+$53 46 24.20& 0.1695& $-$1.05& 23.24 & $-$21.52 &   \nodata           				 \\ 
1262111&11 35 26.68&$+$31 53 33.14& 0.2310& $-$1.05& 24.71 & $-$22.67 & {\tt MACS J1135.4+3153}\\
1268493&11 37 50.23&$+$46 36 33.65& 0.3151& $-$1.03& 24.72 & $-$23.08 &  \nodata                                          \\
1286560&11 44 27.21&$+$37 08 32.42& 0.1148& $-$1.56& 23.55 & $-$21.11 &  \nodata                                           \\
1294316&11 47 12.35&$+$38 19 26.32& 0.5977& $-$1.01& 25.34 & $-$22.11 &  \nodata                                           \\
1304138&11 50 49.21&$+$62 19 49.04& 0.3453& $-$1.69& 24.37 & $-$23.04 &  \nodata                                           \\
1307368&11 51 58.63&$+$31 40 32.05& 0.5079& $-$1.04& 25.69 & $-$23.39 &  \nodata                                          \\
1307436&11 52 00.09&$+$33 13 42.49& 0.3573& $-$1.48& 23.97 & $-$23.15 & \nodata \\
\hline
\end{tabular}
\end{small}
\end{center}
\end{table*}

\small
\begin{table*}
{\bf Table 1.} \\
\begin{center}
\begin{small}
\begin{tabular}{rrrrrcrc}
\hline
ID & R.A$^{radio}_{J2000}$ & DEC$^{radio}_{J2000}$ & z & $\alpha_{352}^{1400}$ & Log($L_{1.4}$)  &$M_r$  & ID from literature\\
& $^h\;\; ^m\;\;\;\; ^s\;\;\,$ & \degr$\;\;\;$ \arcmin$\;\;\;$ \arcsec$\;$ && &W Hz$^{-1}$ & &Designation\\
\hline
1309157&11 52 36.33&$+$37 32 43.86& 0.2294& $-$1.19& 24.22 & $-$20.23  &{\tt MaxBCG J178.15191+37.54548  }                 \\
1319027&11 56 05.51&$+$34 33 05.33& 0.2536& $-$1.10& 25.05 & $-$21.83  &{\tt [EAD2007] 200  Arcs     }                     \\ 
1326519&11 58 48.07&$+$57 17 19.11& 0.2598& $-$1.04& 25.12 & $-$24.49  & \nodata                                           \\ 
1348701&12 06 47.88&$+$51 57 10.95& 0.3446& $-$1.28& 24.66 & $-$21.51  &  \nodata                                         \\ 
1351855&12 08 00.78&$+$43 39 19.12& 0.2657& $-$1.00& 24.77 & $-$22.57  &{\tt MaxBCG J182.00318+43.65537 }                  \\ 
1353531&12 08 37.16&$+$61 21 06.52& 0.2748& $-$1.48& 23.96 & $-$22.17   & \nodata                                          \\
1354974&12 09 08.84&$+$44 00 11.30& 0.0376& $-$1.12& 22.12 & $-$22.98  &{\tt NGC4135, (G. group)}                                      \\ 
1362044&12 11 46.22&$+$32 38 38.16& 0.6115& $-$1.07& 24.94 & $-$22.25  & \nodata             				  \\ 
1406400&12 28 02.17&$+$34 40 40.12& 0.2775& $-$1.40& 23.81 & $-$22.53  &{\tt MaxBCG J187.00902+34.67753}                   \\ 
1439302&12 40 04.88&$+$37 44 15.46& 0.1879& $-$1.15& 23.75 & $-$21.53  & {\tt NSC J124001+374544}             				  \\ 
1468909&12 51 07.51&$+$56 25 44.98& 0.2008& $-$1.22& 23.42 & $-$21.73  &  \nodata            				  \\ 
1505383&13 04 31.36&$+$51 43 42.64& 0.2757& $-$1.56& 24.43 & $-$22.28  &{\tt MaxBCG J196.15441+51.71551}\\
1510127&13 06 12.17&$+$51 44 06.94& 0.2773& $-$1.16& 24.55 & $-$22.54  &{\tt MaxBCG J196.55069+51.73530}\\
1532575&13 14 18.32&$+$41 24 30.18& 0.1987& $-$1.05& 23.25 & $-$20.90  &{\tt MaxBCG J198.57609+41.40825}\\
1547051&13 19 38.92&$+$61 39 11.68& 0.1333& $-$1.21& 23.49 & $-$22.47  &  \nodata            				   \\
1559285&13 24 12.38&$+$31 17 24.33& 0.4268& $-$1.19& 24.63 & $-$22.47  &  \nodata            				   \\
1604023&13 40 32.89&$+$40 17 38.79& 0.1719& $-$1.17& 23.32 & $-$22.35  &{\tt RX J1340.5+4017 GGroup}\\
1607910&13 41 59.68&$+$42 21 32.32& 0.4261& $-$1.22& 24.07 & $-$22.75   & \nodata            				   \\
1627288&13 49 03.74&$+$30 52 27.51& 0.0814& $-$1.13& 22.53 & $-$19.43   & \nodata           				   \\
1628350&13 49 27.88&$+$46 20 15.29& 0.4212& $-$1.14& 24.17 & $-$14.98   & \nodata            				   \\
1673236&14 06 03.34&$+$52 09 51.98& 0.4823& $-$1.01& 24.75 & $-$23.38   & \nodata           				   \\
1707796&14 18 37.62&$+$37 46 22.63& 0.1349& $-$1.34& 23.37 & $-$21.87  &{\tt ABELL 1896}\\
1714152&14 20 56.84&$+$53 13 07.25& 0.7430& $-$1.05& 25.29 & $-$23.57   & \nodata                                            \\
1714768&14 21 10.18&$+$42 09 12.97& 0.3529& $-$1.12& 24.03 & $-$20.99  &{\tt NSCS J142115+420743}\\
1735168&14 28 41.23&$+$43 41 34.03& 0.2136& $-$1.04& 23.70 & $-$22.42  &{\tt NSCS J142842+434009}\\
1735918&14 28 57.67&$+$54 36 27.65& 0.3819& $-$1.50& 24.39 & $-$21.60    &             				   \\
1744373&14 32 04.05&$+$46 37 43.79& 0.0927& $-$1.03& 22.65 & $-$19.72    &{\tt NSC J143143+463738}\\
1755382&14 36 02.53&$+$33 07 53.79& 0.0939& $-$1.02& 22.80 & $-$20.51    &  \nodata         				  \\ 
1756038&14 36 19.44&$+$48 32 10.68& 0.1912& $-$1.07& 24.03 & $-$21.62    &  \nodata            				  \\
1759889&14 37 42.41&$+$39 27 45.12& 0.2455& $-$1.36& 24.35 & $-$21.94   &{\tt MaxBCG J219.42655+39.46313} \\   
1768885&14 40 57.03&$+$46 36 46.91& 0.8395& $-$1.12& 25.35 & $-$19.00    &   \nodata       				\\  
1775379&14 43 17.07&$+$46 43 48.40& 0.2424& $-$1.34& 23.68 & $-$21.83    &   \nodata        				\\   
1793973&14 50 03.51&$+$31 30 15.02& 0.2746& $-$1.10& 24.25 & $-$21.95    & {\tt MaxBCG J222.55394+31.49750}\\
1795282&14 50 31.54&$+$32 53 03.73& 0.1775& $-$1.38& 23.45 & $-$21.67    &   \nodata       				\\    
1830754&15 03 23.78&$+$46 06 16.28& 0.4269& $-$1.06& 24.19 & $-$22.74    &   \nodata       				\\    
1836180&15 05 23.43&$+$47 06 25.59& 0.2615& $-$1.63& 23.90 & $-$22.52   &{\tt ABELL 2024} \\    
1837171&15 05 46.23&$+$54 54 01.56& 0.2824& $-$1.24& 24.49 & $-$22.69    &   \nodata         				\\    
1838173&15 06 08.41&$+$60 02 16.86& 0.5196& $-$1.03& 24.72 & $-$24.11    &   \nodata         				\\    
1862173&15 15 05.54&$+$43 09 01.38& 0.0177& $-$1.14& 21.11 & $-$23.69     & {\tt CGCG 221-045}        				\\    
1929427&15 39 50.77&$+$30 43 03.90& 0.0971& $-$1.18& 23.10 & $-$22.36    &{\tt MaxBCG J234.96158+30.71777}  \\     
1932815&15 41 05.46&$+$32 04 50.85& 0.0529& $-$1.06& 22.51 & $-$20.79    &   \nodata        				\\    
1934628&15 41 46.53&$+$45 56 14.29& 0.2024& $-$1.08& 24.31 & $-$21.37    &   \nodata         				\\    
1958800&15 50 51.44&$+$42 02 30.47& 0.0334& $-$1.09& 21.91 & $-$20.61     &  \nodata \\
1963482&15 52 41.11&$+$37 24 34.16& 0.3710& $-$1.92& 24.23 & $-$23.69     &  \nodata         				\\    
1967858&15 54 23.54&$+$48 41 07.36& 0.2271& $-$1.04& 23.93 & $-$22.32    &{\tt MaxBCG J238.59817+48.68496} \\    
2001242&16 07 25.43&$+$47 50 24.15& 0.3282& $-$1.02& 24.76 & $-$22.54  &{\tt ABELL 2157}\\
2064005&16 33 10.92&$+$36 07 35.15& 0.1648& $-$1.02& 23.93 & $-$21.42  &{\tt  MaxBCG J248.29532+36.12611} \\    
2088195&16 43 26.82&$+$39 30 39.92& 0.4119& $-$1.05& 24.72 & $-$22.15   &    \nodata          		\\   
2122675&16 59 01.01&$+$32 29 38.93& 0.0627& $-$1.26& 23.82 & $-$21.61  &{\tt ABELL 2241}\\
2134416&17 04 26.40&$+$39 10 12.25& 0.1283& $-$1.01& 23.23 & $-$21.34   &{\tt NSC J170432+390956  }            \\    
\hline
\end{tabular}
\end{small}
\end{center}
\label{table}
\end{table*}

\section{Acknowledgments}
We are grateful to the anonymous referee for his/her careful reading of the manuscript and a number of comments, which improved the paper.
This work was partially supported by the Consejo Nacional de Investigaciones Cient\'{\i}ficas y T\'ecnicas (CONICET),the Secretar\'ia de Ciencia y T\'ecnica de la Universidad Nacional de C\'ordoba . The authors made use of the database CATS (Verkhodanov et al. 1997) of the Special Astrophysical Observatory and the NASA/IPAC extragalactic database (NED) which is operated by the Jet Propulsion Laboratory, Caltech, under contract with the National Aeronautics and Space Administration. 
This publication makes use of data products from the Sloan Digital Sky Survey (SDSS). Funding
for the SDSS and SDSS-VI has been provided by the Alfred P. Sloan
Foundation, the Participating Institutions, the National Science
Foundation, the U.S. Department of Energy, the National Aeronautics
and Space Administration, the Japanese Monbukagakusho, the
Max Planck Society, and the Higher Education Funding Council for
England. The SDSS Web Site is http://www.sdss.org/. The SDSS is
managed by the Astrophysical Research Consortium for the Participating
Institutions. The Participating Institutions are the American
Museum of Natural History, Astrophysical Institute Potsdam, University
of Basel, University of Cambridge, Case Western Reserve
University, University of Chicago, Drexel University, Fermilab, the
Institute for Advanced Study, the Japan Participation Group, Johns
Hopkins University, the Joint Institute for Nuclear Astrophysics,
the Kavli Institute for Particle Astrophysics and Cosmology, the
Korean Scientist Group, the Chinese Academy of Sciences (LAMOST),
Los Alamos National Laboratory, the Max-Planck-Institute
for Astronomy (MPIA), the Max-Planck-Institute for Astrophysics
(MPA), New Mexico State University, Ohio State University, University
of Pittsburgh, University of Portsmouth, Princeton University,
the United States Naval Observatory, and the University of
Washington.
{}


\end{document}